\documentclass{nature}
\usepackage[utf8]{inputenc}
\usepackage{graphicx}
\graphicspath{ {./figures/} }
\usepackage{textcomp,gensymb}
\usepackage[euler]{textgreek}
\usepackage{lineno}
\usepackage{caption}
\DeclareCaptionLabelFormat{vertline}{#1 #2 $|$}
\usepackage{prettyref}
\newrefformat{fig}{Fig.~\ref{#1}}
\newrefformat{exfig}{Extended Data Fig.~\ref{#1}}

\usepackage[normalem]{ulem}

\usepackage{color}

\usepackage{units}
\usepackage{amsmath}
\usepackage{comment}

\makeatletter
\let\saved@includegraphics\includegraphics
\AtBeginDocument{\let\includegraphics\saved@includegraphics}
\renewenvironment{figure}{\@float{figure}}{\end@float}

\makeatother
\newcommand{\beginsupplement}{%
        \setcounter{table}{0}
        \renewcommand{\table}{\arabic{table}|}%
        \setcounter{figure}{0}
        \renewcommand{\figurename}{Extended Data Figure}
     }

\title{Superconductivity without insulating states in twisted bilayer graphene stabilized by 
monolayer WSe\textsubscript{2}}

\author{Harpreet Singh Arora$^{1,2*}$, Robert Polski$^{1,2*}$, Yiran Zhang$^{1,2,3*}$, Alex Thomson$^{2,3,4}$, 
Youngjoon Choi$^{1,2,3}$, Hyunjin Kim$^{1,2,3}$,  Zhong Lin$^5$, Ilham Zaky Wilson$^5$, Xiaodong Xu$^{5,6}$, Jiun-Haw Chu$^5$, 
Kenji Watanabe$^7$, Takashi Taniguchi$^7$, Jason Alicea$^{2,3,4}$ and Stevan Nadj-Perge$^{1,2\dagger}$}

\begin{document}

\maketitle

\begin{affiliations}

    \item T. J. Watson Laboratory of Applied Physics, California Institute of Technology, 
         1200 East California Boulevard, Pasadena, California 91125, USA
    \item Institute for Quantum Information and Matter, California Institute of Technology, Pasadena, California 91125, USA
    \item Department of Physics, California Institute of Technology, Pasadena, California 91125, USA
    \item Walter Burke Institute for Theoretical Physics, California Institute of Technology, Pasadena, California 91125, USA
    \item Department of Physics, University of Washington, Seattle, Washington 98195, USA
    \item Department of Materials Science and Engineering, University of Washington, Seattle, WA
98195, USA
    \item National Institute for Materials Science, Namiki 1-1, Tsukuba, Ibaraki 305 0044, Japan
    \item[*] These authors contributed equally to this work
    \item[$^\dagger$] Correspondence: s.nadj-perge@caltech.edu 
\end{affiliations}

%\begin{linenumbers}

\begin{abstract}
Magic-angle twisted bilayer graphene (TBG), with rotational misalignment close to 1.1\degree, 
features isolated flat electronic bands that host a rich phase diagram of correlated 
insulating, superconducting, ferromagnetic, and topological 
phases\cite{caoCorrelatedInsulatorBehaviour2018, 
caoUnconventionalSuperconductivityMagicangle2018,
yankowitzTuningSuperconductivityTwisted2019,
sharpeEmergentFerromagnetismThreequarters2019, 
luSuperconductorsOrbitalMagnets2019,
serlinIntrinsicQuantizedAnomalous2019}. 
The origins of the correlated insulators and superconductivity, and the interplay between them, 
are particularly elusive due to the sensitivity of these correlated 
states to microscopic details. 
Both states have been previously observed only for angles within $\pm$0.1\degree~from 
the magic-angle value and occur in adjacent or overlapping electron density ranges; nevertheless, 
it is still unclear how the two states are related.
Beyond the twist angle and strain, the dependence of the TBG phase diagram on the 
alignment\cite{sharpeEmergentFerromagnetismThreequarters2019,serlinIntrinsicQuantizedAnomalous2019} 
and thickness of insulating hexagonal boron nitride 
(hBN)\cite{stepanovInterplayInsulatingSuperconducting2019, saitoDecouplingSuperconductivityCorrelated2019a} 
used to encapsulate the graphene sheets indicates the importance of the microscopic 
dielectric environment.  Here we show that adding an 
insulating tungsten-diselenide (WSe$_2$) monolayer between hBN and TBG 
stabilizes superconductivity at twist angles much smaller than the established magic-angle value.
For the smallest angle of $\mathbf{\theta}$ = 0.79\degree, we still 
observe clear superconducting signatures, despite the complete absence of the correlated 
insulating states and vanishing gaps between the dispersive and flat bands. 
These observations demonstrate that, even though electron correlations may be 
important, superconductivity in TBG can exist even when TBG exhibits metallic behaviour 
across the whole range of electron density. Finite-magnetic-field measurements 
further reveal breaking of the four-fold spin-valley symmetry in the system, consistent with 
large spin-orbit coupling induced in TBG via proximity to WSe$_2$. The survival of 
superconductivity in the presence of spin-orbit coupling imposes additional constraints 
on the likely pairing channels. Our results highlight the importance of symmetry 
breaking effects in stabilizing electronic states in TBG and open new avenues for 
engineering quantum phases in moir\'e systems. 
\end{abstract}

%==============
Strongly correlated electron systems often exhibit a variety of 
quantum phases with similar ground-state energies, separated by phase boundaries that 
depend sensitively on microscopic details. Twisted bilayer graphene, with 
twist angle close to the magic angle $\theta$\textsubscript{M} $\approx$ 1.1\degree, has recently emerged 
as a highly tunable platform with an exceptionally rich phase 
diagram\cite{caoCorrelatedInsulatorBehaviour2018,caoUnconventionalSuperconductivityMagicangle2018} 
hosting correlated insulating states, superconductivity, and 
ferromagnetism\cite{yankowitzTuningSuperconductivityTwisted2019,
sharpeEmergentFerromagnetismThreequarters2019, 
luSuperconductorsOrbitalMagnets2019,
serlinIntrinsicQuantizedAnomalous2019}. Strong correlations in 
TBG originate from the non-dispersive (flat) bands that are created by the hybridization of 
the graphene sheets\cite{bistritzerMoireBandsTwisted2011} 
and are isolated 
from the rest of the energy spectrum by an energy gap 
$\sim$ 30 meV\cite{choiElectronicCorrelationsTwisted2019, polshynLargeLinearintemperatureResistivity2019}. 
Previous transport experiments on magic-angle TBG found that the correlated insulators are 
often accompanied by superconductivity in a narrow range $\pm$0.1\degree\, around 
$\theta$\textsubscript{M}\cite{caoCorrelatedInsulatorBehaviour2018,yankowitzTuningSuperconductivityTwisted2019, saitoDecouplingSuperconductivityCorrelated2019a, stepanovInterplayInsulatingSuperconducting2019, caoStrangeMetalMagicangle2019}, 
with signatures of these states observed down 
to 0.93\degree \cite{codecidoCorrelatedInsulatingSuperconducting2019}. Close to 
$\theta$\textsubscript{M}, the correlated insulators develop at electron densities 
that correspond to an integer number $\nu$ of electrons per moir\' e unit 
and are surrounded by intermittent pockets of 
superconductivity\cite{luSuperconductorsOrbitalMagnets2019}; 
both phases appear most frequently around 
$\nu$ = $\pm$2\cite{caoUnconventionalSuperconductivityMagicangle2018,yankowitzTuningSuperconductivityTwisted2019}. 
Away from $\theta$\textsubscript{M}, however, both phases are suppressed 
as the effects of electron-electron interactions quickly diminish due 
to a rapid increase of  
the flat-band bandwidth and corresponding dominance of kinetic energy\cite{bistritzerMoireBandsTwisted2011,caoCorrelatedInsulatorBehaviour2018}. 
In addition to the TBG twist angle, the physics of the correlated phases is also affected by the hBN employed 
as a high-quality dielectric. In particular, since hBN and graphene exhibit similar crystal 
lattices, the relative alignment between the hBN and TBG is critical. For example,
a ferromagnetic state near $\nu$ = +3 was observed in devices where hBN aligns with TBG\cite{sharpeEmergentFerromagnetismThreequarters2019, serlinIntrinsicQuantizedAnomalous2019}. 
However, in such devices the band structure of the flat bands is strongly 
altered\cite{serlinIntrinsicQuantizedAnomalous2019}, and superconductivity---typically observed 
when hBN and TBG are misaligned---is absent. Recent work using a very thin hBN layer 
separating a back gate from TBG 
additionally suggests that electrostatic screening plays a prominent role in the appearance of insulating and 
superconducting states\cite{stepanovInterplayInsulatingSuperconducting2019}. These experiments 
exemplify the effects of hBN layers on the phase diagram in hBN-TBG-hBN structures and highlight the 
importance of understanding how microscopic details of the dielectric environment 
alter the properties of correlated phases.

\begin{figure}
\captionsetup{format=plain,labelsep=space}
\begin{center}
    \includegraphics[width=15cm]{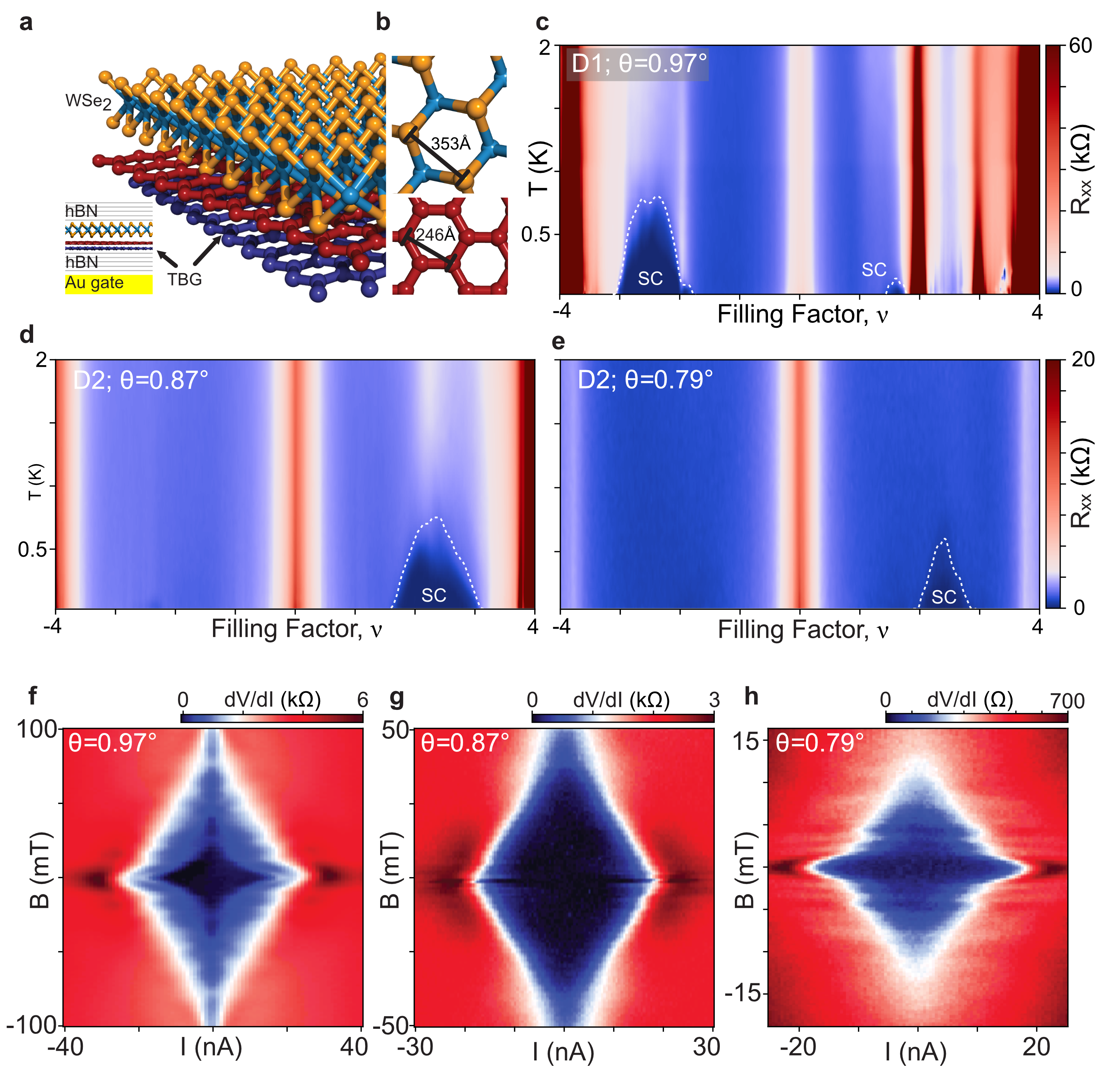}
\end{center}
\caption{{\bf Superconductivity in small-angle TBG-WSe\textsubscript{2} structures.} {\bf a}, 
Schematic of the TBG-WSe\textsubscript{2} structure showing the crystal lattice of two 
graphene layers (red and blue) and WSe\textsubscript{2} (yellow and cyan). Inset: Complete structure 
including encapsulating hBN layers on top and bottom and a gold back-gate. {\bf b}, Top view 
of WSe\textsubscript{2} and graphene, indicating different unit-cell sizes. 
{\bf c-e}, Longitudinal resistance R\textsubscript{xx} vs.~temperature and electron density, expressed as 
a flat-band filling factor \textnu, for devices D1 and D2 and angles 
$\theta$ = 0.97\degree, {\bf c}; $\theta$ = 0.87\degree, {\bf d}; 
and $\theta$ = 0.79\degree, {\bf e}. In device D2, adjacent sets of electrodes have 
slightly different twist angle, as explained in the Methods section. 
Superconducting domes (SC) are indicated by a dashed line that delineates half of the resistance 
measured at 2 K (except for the electron-side dome for 0.97\degree, for which the normal temperature used was taken at 1K). 
{\bf f-h}, Fraunhofer-like interference patterns, typically observed 
in TBG superconducting devices, for the three contact pairs 
($\theta$ = 0.97\degree, $\nu$ = --2.40, % VBg=-4.2V  
{\bf f}; $\theta$ = 0.87\degree, $\nu$ = 1.96 %Vbg = 2.3 V
{\bf g}; and $\theta$ = 0.79\degree, $\nu$ = 2.30, %Vbg = 2.3 V 
{\bf h}). }
\label{fig:tbgSC}
\end{figure}

%=======================
Here, instead of the usual hBN-TBG-hBN structures, we investigate devices made from 
hBN-TBG-WSe\textsubscript{2}-hBN van der Waals stacks in which a monolayer of WSe\textsubscript{2} 
resides between the top hBN and TBG (\prettyref{fig:tbgSC}a). Our stacks are assembled using a 
modified ‘tear and stack’ technique where the ‘tearing’ and ‘stacking’ of TBG is facilitated by 
monolayer WSe\textsubscript{2}; see Methods and Extended Data Fig. 1 %\prettyref{exfig:fabrication} 
for fabrication 
details. Like hBN, flakes of transition metal dichalcogenides, such as WSe\textsubscript{2}, 
can be used as a high-quality insulating dielectric for graphene-based 
devices\cite{kretininElectronicPropertiesGraphene2014}; however, the two van der Waals dielectrics  
differ in several ways that may alter the TBG band structure. First, unlike hBN, the WSe$_2$ and 
graphene lattice constants differ significantly (0.353 nm for WSe\textsubscript{2} 
and 0.246 nm for graphene, \prettyref{fig:tbgSC}b). This mismatch implies that the moir\' e pattern formed between 
TBG and WSe\textsubscript{2} has a maximum lattice constant  $\sim$ 1 nm, in other words, much smaller than that formed 
in small-angle TBG ($>$ 10 nm). Second, it is well-established that WSe\textsubscript{2} can induce a spin-orbit 
interaction (SOI) in graphene via van der Waals 
proximity\cite{islandSpinOrbitdrivenBand2019,wangQuantumHallEffect2019}. And finally, 
due to hybridization effects, WSe\textsubscript{2} may also change both the Fermi velocity of the 
proximitized graphene sheet and the system's phonon spectrum. We chose to use monolayer WSe\textsubscript{2} 
in particular because of its large band 
gap\cite{wilsonDeterminationBandOffsets2017} that allows applying a large range of gate voltages. 
It has also been suggested previously that a monolayer 
induces larger spin-orbit coupling in graphene compared to a few-layer 
WSe\textsubscript{2}\cite{wakamuraSpinorbitInteractionInduced2019}.

%######################################################
We have studied three TBG-WSe$_2$ devices and show results for two of them
in the main text (see Extended Data Fig. 4
%\prettyref{exfig:D3extra} 
for data from an additional device). 
Surprisingly, we find robust superconductivity in all studied TBG-WSe\textsubscript{2} structures, even 
for twist angles far outside of the previously established range. \prettyref{fig:tbgSC}c-e shows
the temperature dependence of resistance over three TBG regions corresponding to angles  
$\theta$ = 0.97\degree, $\theta$ = 0.87\degree\, and $\theta$ = 0.79\degree; in all cases 
superconducting transitions are clearly visible. Aside from the drop in R\textsubscript{xx} to zero, 
we also observe well-resolved Fraunhofer-like patterns for all three angles (Figs. 1f-1h), qualitatively 
similar to the typical hBN-TBG-hBN devices\cite{caoUnconventionalSuperconductivityMagicangle2018,luSuperconductorsOrbitalMagnets2019,
yankowitzTuningSuperconductivityTwisted2019}. The small periodic modulations 
of the critical current in magnetic field have been previously attributed to the presence of 
Josephson junctions in the system, independently corroborating the presence of superconducting 
correlations. In our devices, we typically see periods of 1.5-3 mT that, if interpreted as the effective 
junction area $S\sim$ 0.67--1.33 \textmu m\textsuperscript{2}, are consistent with the device geometry.  

\begin{figure}
\begin{center}
    \includegraphics[width=16cm]{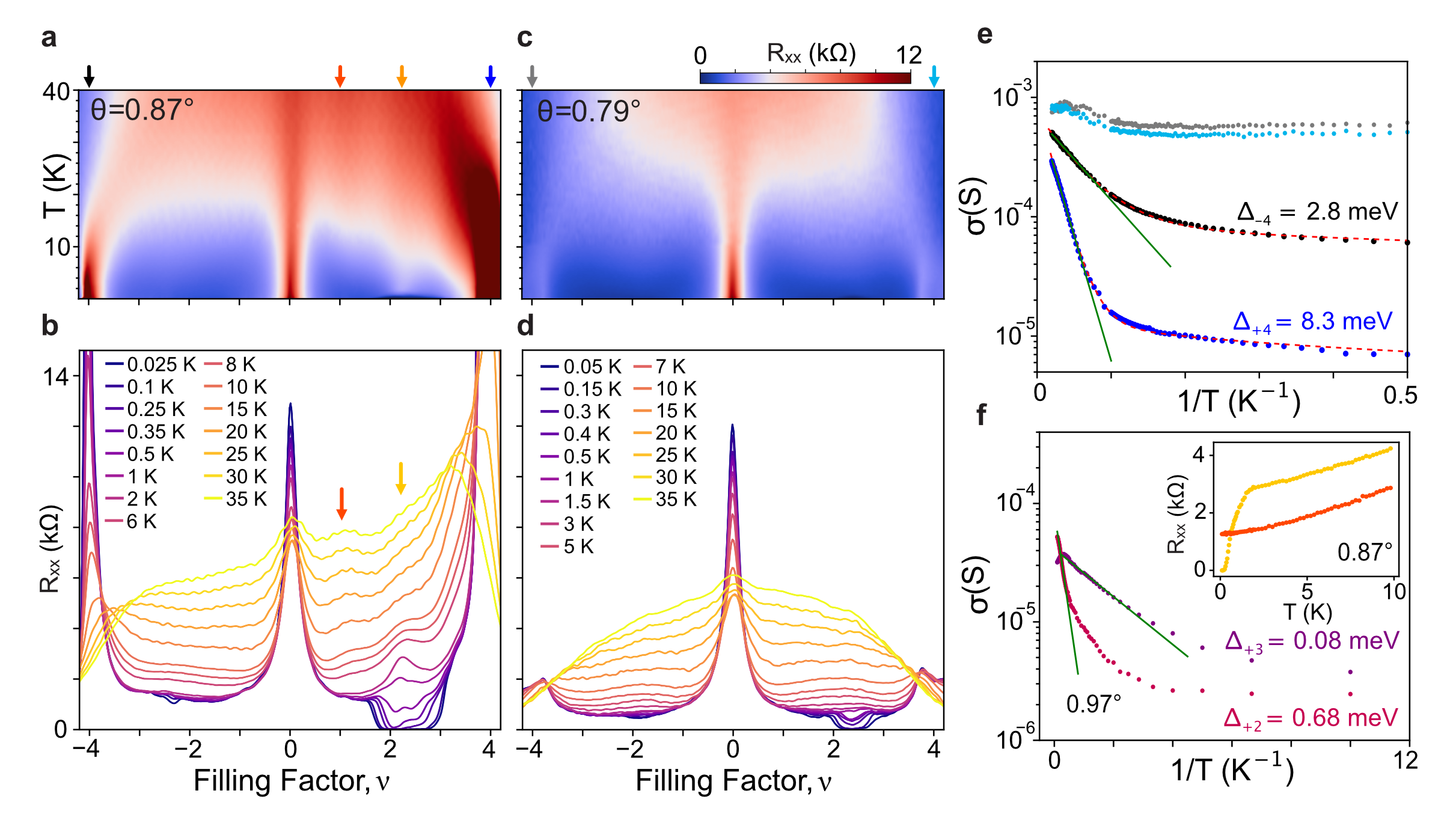}
\end{center}
\caption{{\bf Absence of correlated insulating states and diminished gap between  
flat and dispersive bands.} {\bf a-d}, Larger-temperature-range data showing R\textsubscript{xx} 
as a function of filling factor \textnu\, for 
$\theta$ = 0.87\degree\, ({\bf a},{\bf b}) and $\theta$ = 0.79\degree\, ({\bf c},{\bf d}).
Line cuts shown in {\bf b} ({\bf d}) are taken from the same data set as {\bf a} ({\bf c}). {\bf e}, 
Conductance vs. 1/T for full filling $\nu$ = $\pm$4 extracted from the data in {\bf a} (blue and black) 
and {\bf c} (cyan and gray). Green and red lines are fits for $\theta$ = 0.87\degree\, to a model that 
includes only activation (green) and both activation and variable-range 
hopping of the form  $\text{exp}[-(T_0/T)^{1/3}]$ \cite{caoSuperlatticeInducedInsulatingStates2016} (red). The gap values shown 
are extracted from the activation only fits (to the form of $\sigma \propto e^{-\mathrm{\Delta}/2k_B T}$); 
the more complete model gives similar gap values of $\mathrm{\Delta_{+4}}$ = 9.4 meV and $\mathrm{\Delta_{-4}}$ = 3.7 meV. 
The behavior for $\theta$ = 0.79\degree~
shows much smaller variation in temperature. 
{\bf f}, Conductance vs. 1/T for partial filling factors $\nu$ = $\pm$2 for 0.97\degree~showing insulating behavior. 
In contrast, for 0.87\degree~partial fillings $\nu$ = $\pm$2 show metallic behavior, see inset.}
\label{fig:hightempgaps}
\end{figure}

For the largest angle $\theta$ = 0.97\degree, a superconducting pocket 
emerges on the hole side near $\nu$ = --2 with a maximal transition temperature 
T\textsubscript{c} $\approx$ 0.8 K.  To our knowledge, this already is the smallest angle for which superconductivity 
has been observed for hole doping. Careful inspection reveals another weak superconductivity pocket
close to $\nu$ = +2 (the behavior at low fields is displayed in Extended Data Fig. 6). 
%\prettyref{exfig:D1fraunhofer}). 
However, despite the small twist angle---falling 
outside the $\theta$\textsubscript{M} $\pm$ 0.1\degree\, range---the observed phase diagram resembles that of regular high-quality 
magic-angle hBN-TBG-hBN 
structures\cite{caoUnconventionalSuperconductivityMagicangle2018,luSuperconductorsOrbitalMagnets2019}. 
For this angle, correlated insulating states are also observed for filling factors 
$\nu$ = +2, +3 with activation gaps of $\mathrm{\Delta_{+2}}$ = 0.68 meV and
$\mathrm{\Delta_{+3}}$ = 0.08 meV, whereas at other filling factors correlated states are less 
developed and do not show insulting behavior (see \prettyref{fig:hightempgaps}f and  Extended Data Fig. 3).
%\prettyref{exfig:D1extratemp}). 
%####### 

Although superconductivity persists for all three angles, the correlated insulators 
are quickly suppressed as the twist angle is reduced. This suppression is not surprising, 
as for angles below $\theta$\textsubscript{M}, the bandwidth increases rapidly and, moreover, the characteristic 
correlation energy scale $e^2/4\pi\epsilon L_m$ also diminishes due to an increase in the moir\' e 
periodicity $L_m$ = $a/sin(\theta/2)$ ($a$ = 0.246 nm denotes the graphene lattice 
constant)\cite{bistritzerMoireBandsTwisted2011,caoCorrelatedInsulatorBehaviour2018, kerelskyMaximizedElectronInteractions2019, choiElectronicCorrelationsTwisted2019, xieSpectroscopicSignaturesManybody2019, jiangChargeOrderBroken2019a}.
For the lower angle of $\theta$ = 0.87\degree\, correlated-insulating behavior is heavily 
suppressed at all filling factors.  In \prettyref{fig:tbgSC}d a peak in longitudinal resistance versus 
density is visible only around $\nu$ = +2 above the superconducting transition 
($T_c$ = 600--800 mK). Data for a larger temperature range (\prettyref{fig:hightempgaps}a-b) shows that the resistance 
peak near $\nu$ = +2 survives up to $T$ = 30 K, and also reveals a new peak near $\nu$ = +1 in 
the temperature range 10-35 K. These observations suggest that  electron correlations remain strong, though the corresponding states appear to be metallic as the overall resistance 
increases with temperature. For this angle, we measure activation gaps at full filling (i.e., at $\nu$ = $\pm4$) of $\mathrm{\Delta_{+4}}$ = 8.3 meV and
$\mathrm{\Delta_{-4}}$ = 2.8 meV (\prettyref{fig:hightempgaps}e) ---far smaller 
than the gaps around $\theta_{\mathrm{M}}$, in line with previous results that 
report a disappearance of the band gap separating dispersive and flat bands at around 
$\theta$ = 0.8\degree\cite{koshinoMaximallyLocalizedWannier2018,polshynLargeLinearintemperatureResistivity2019}.

%==========================================
At the smallest angle, $\theta$ = 0.79\degree, along with the lack of insulating states at any partial filling, the resistance at full filling is even more reduced (\prettyref{fig:hightempgaps}c-d). The relatively low resistances 
\textless 2 k\textOmega\, measured at full filling---which are less than $15 \%$ of the resistance at the 
charge neutrality point (CNP)---suggest a semi-metallic band structure around full filling, consistent with theoretical
expectations for TBG at $\theta$ = 0.79\degree \cite{koshinoMaximallyLocalizedWannier2018} and the resistivity of a dilute 2D electron gas\cite{lillyResistivityDilute2D2003}. We emphasize, however,
that despite the absence of  both full-filling band gaps and correlated insulators, the superconducting low-resistance pocket near $\nu=+2$ is clearly resolved 
(Figs.~1e and 1h).
%========================================

Both the disappearance of the correlated insulators and the vanishing gap between flat and dispersive bands for low angles suggest that the additional WSe\textsubscript{2} monolayer does not significantly change the magic 
angle. Since superconductivity survives to much lower angles compared to correlated 
insulating states, the two phenomena appear to have 
different 
origins\cite{stepanovInterplayInsulatingSuperconducting2019, saitoDecouplingSuperconductivityCorrelated2019a}. 
Note also that the close proximity of the dispersive bands does not seem to have a major impact on the 
superconducting phase. While these findings are not consistent with a 
scenario wherein superconductivity descends from a Mott-like insulating state as in high-T\textsubscript{c} 
superconductors\cite{leeDopingMottInsulator2006}, we do emphasize that electron correlations may  
still prove essential for the development of superconductivity. 
For instance, even for the smallest 
angle of $\theta$ = 0.79\degree, the superconducting pocket is seemingly pinned to the vicinity 
of $\nu$ = 2. Additionally, as shown in \prettyref{fig:hightempgaps}, at higher temperatures residual R\textsubscript{xx} peaks  
can still appear at certain integer filling factors despite the absence of gapped correlated insulating states. It is thus 
hard to rule out the possibility that superconductivity arises from correlated states of metallic nature that may be present 
at smaller angles and near integer values of $\nu$ in analogy to other exotic superconducting 
systems\cite{stewartHeavyfermionSystems1984, ardavanRecentTopicsOrganic2011, stewartSuperconductivityIronCompounds2011}.

\begin{figure}
\begin{center}
    \includegraphics[width=16.5cm]{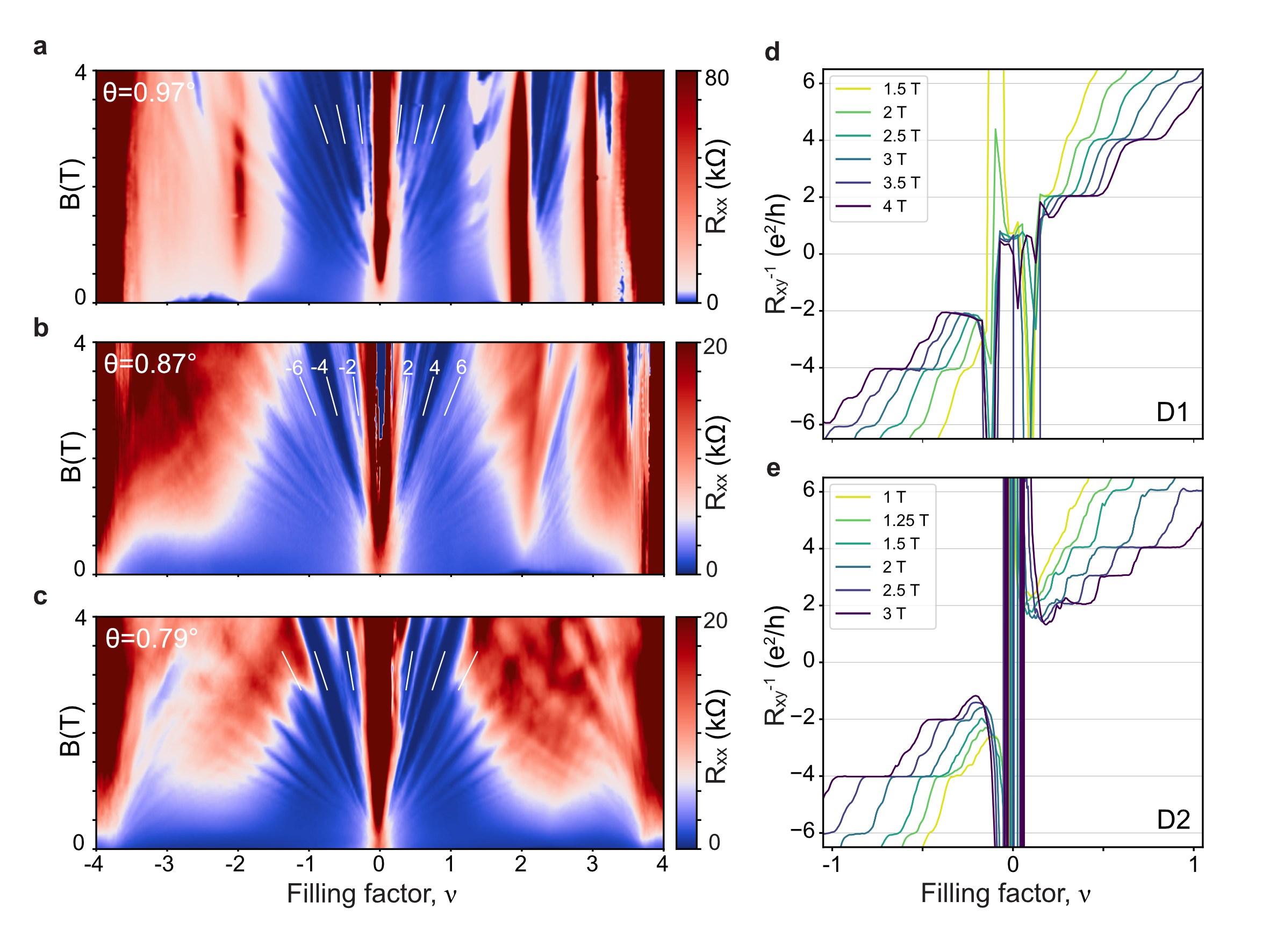}
\end{center}
\caption{{\bf Breaking of the four-fold degeneracy.} {\bf a-c}, Longitudinal resistance R\textsubscript{xx} as a function 
of magnetic field and \textnu ~for the three contacts pairs yielding $\theta$ = 0.97\degree, $\theta$ = 0.87\degree, and $\theta$ = 0.79\degree.
In contrast to typical hBN-TBG-hBN devices, here the dominant sequence in the Landau fan is $\pm$2,$\pm$4, $\pm$6 indicating breaking of the four
fold (spin-valley) symmetry. {\bf d-e}, Hall conductance for devices D1 ($\theta$ = 0.97\degree) and D2 showing quantized steps 
around the CNP with steps corresponding to $\pm$2, $\pm$4 and $\pm$6 being pronounced down to 1.5 T. }
\label{fig:landaufans}
\end{figure}

%============================================
Measurements in finite magnetic field reveal further insights into the physics of TBG-WSe\textsubscript{2} 
structures (\prettyref{fig:landaufans}). Surprisingly, for all three angles we find that even at modest magnetic fields, above 
$B$ = 1 T, gaps between Landau levels are well-resolved, showing a fan diagram that diverges from the CNP. The slopes of the dominant sequence of  R\textsubscript{xx} minima correspond to even-integer Landau level fillings $\pm$2, $\pm$4, $\pm6$, etc.---indicating broken four-fold (spin-valley flavor) symmetry. By contrast, the majority of previous 
transport 
experiments\cite{caoUnconventionalSuperconductivityMagicangle2018, yankowitzTuningSuperconductivityTwisted2019} 
near the magic angle report a Landau-fan sequence $\pm$4, $\pm$8, $\pm$12 
at the CNP,  with broken-symmetry states being only occasionally observed at 
the lowest Landau level (corresponding to the $\pm$2 sequence)\cite{sharpeEmergentFerromagnetismThreequarters2019,luSuperconductorsOrbitalMagnets2019, uriMappingTwistAngle2019}. 
In addition to R\textsubscript{xx} minima corresponding to the gaps between 
Landau levels, we also measured quantized Hall conductance plateaus, further 
corroborating the two-fold symmetry and indicating the low disorder in the 
measured TBG areas. Note also that for the smallest angle ($\theta$ = 
0.79\degree\,) we do not observe obvious signatures of correlated insulating 
states near $\nu$ = 2 up to $B$ = 4 T. 

\begin{figure}
    \begin{center}
        \includegraphics[width=16cm]{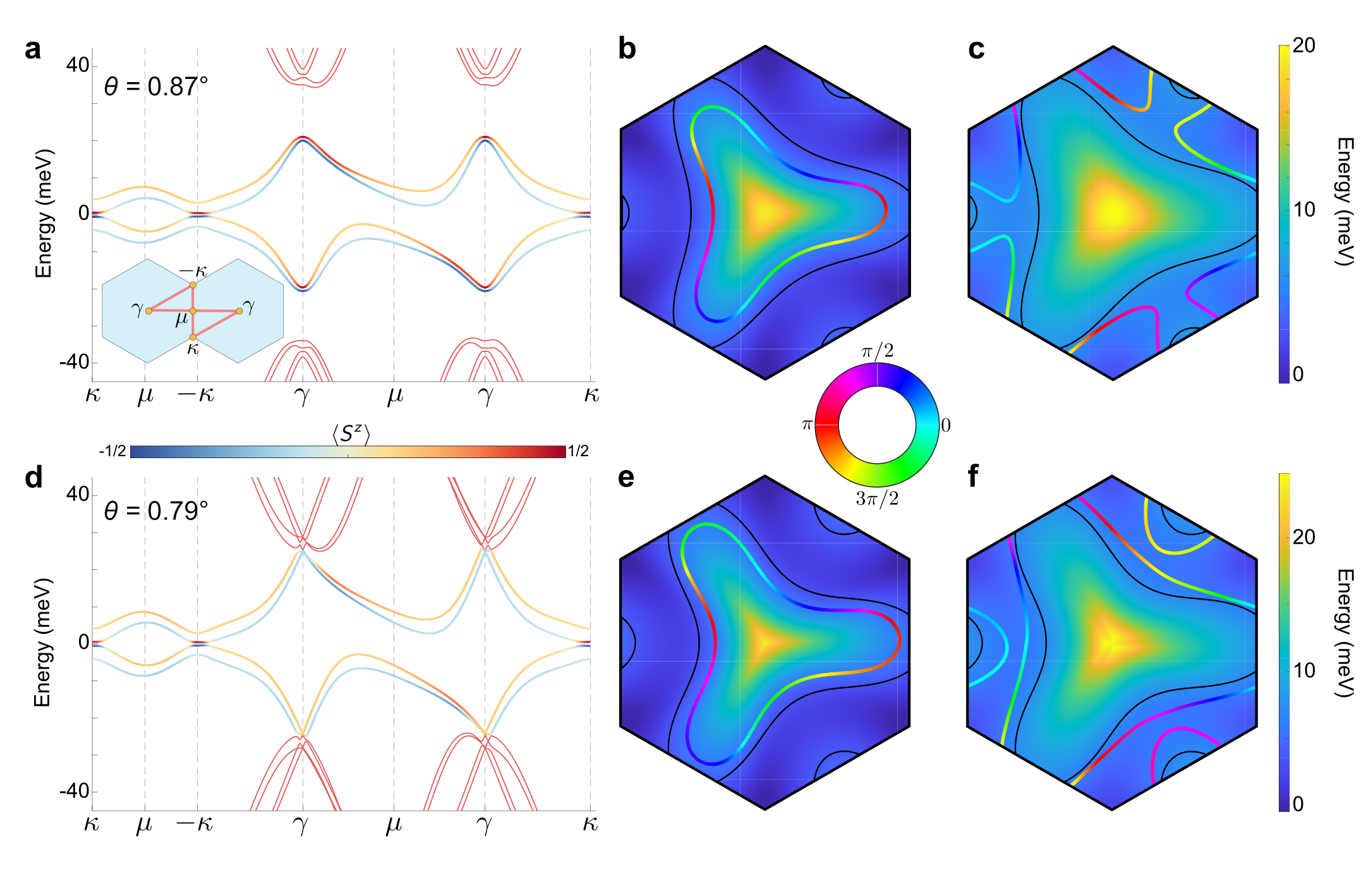}
    \end{center}
    \caption{{\bf Spin-orbit effect on TBG band structure.} Continuum-modeling results for valley $K$ that include Ising and Rashba spin-orbit 
    coupling at 0.87\degree ({\bf a}-{\bf c}) and 0.79\degree ({\bf d}-{\bf f}) twist angles. Details are given in the Methods section.
 ({\bf a, d}) Band structure along high-symmetry directions of the Brillouin zone indicated in the inset.  The line color represents the out-of-plane 
 spin projection, $\langle S_z\rangle $. ({\bf b},{\bf c}) and ({\bf e}, {\bf f}) Energy of the upper pair of flat conduction bands, including spin-orbit 
 coupling.  Colored lines show the Fermi surfaces at $\nu$ = +2, with the color indicating the in-plane spin projection. The out-of-plane projection is 
 largely constant along these surfaces and may therefore be deduced from {\bf a} and {\bf d}. Black lines correspond to the Fermi surface without SOI effects. 
 The large spin-orbit-induced Fermi-surface deformation visible here reflects the flatness of the bands near the Fermi energy. 
 Persistence of superconductivity with such distortions constrains the likely pairing channel in TBG.}
\label{fig:spinorbittheory}
\end{figure}

%============================================
The observed two-fold degeneracy is consistent with a scenario in which the TBG band structure is modified by the 
spin-orbit interaction (SOI) inherited from the WSe\textsubscript{2} monolayer (\prettyref{fig:spinorbittheory}). Previous works established that 
WSe\textsubscript{2} can induce large SOI of both Ising and Rashba type into monolayer and bilayer 
graphene\cite{islandSpinOrbitdrivenBand2019,wangQuantumHallEffect2019}, and it is 
therefore reasonable to assume that the SOI is similarly generated in the upper (proximitized) layer of TBG in our devices. 
Continuum-model calculations taking into account this effect show that the SOI lifts the degeneracy
of both flat and dispersive bands, thereby breaking four-fold spin-valley symmetry. In a finite magnetic field, 
the resulting Landau levels then descend from Kramer's states that are only two-fold degenerate.
Odd steps---which are not generated by the SOI---are occasionally observed for low angles. 
We attribute these steps to additional symmetry breaking, possibly due to correlation 
effects originating either from flat-band physics or simply a magnetic-field-induced effect at 
low electronic densities.

Induced SOI can additionally constrain the nature of the TBG phase diagram.  In particular, the SOI 
acts as an explicit symmetry-breaking field that further promotes instabilities favoring
compatible symmetry-breaking patterns while suppressing those that do not.  The relative
robustness of the $\nu$ = 2 correlated insulator in our $\theta$ = 0.97\degree ~device
suggests that interactions favor re-populating 
bands\cite{zondinerCascadePhaseTransitions2019,wongCascadeTransitionsCorrelated2019}
in a manner that also satisfies the
spin-orbit energy.  
Furthermore, the survival of superconductivity with SOI constrains the plausible pairing 
channels---particularly given the dramatic spin-orbit-induced Fermi-surface deformations 
that occur at $\nu$ = +2  (\prettyref{fig:spinorbittheory}).
Superconductivity in our low-twist-angle devices, for instance, is consistent with Cooper 
pairing of time-reversed partners that remain resonant with SOI.
Thus the stability of candidate insulating and superconducting phases to the SOI provides a
nontrivial constraint for theory\cite{guineaElectrostaticEffectsBand2018,youSuperconductivityValleyFluctuations2019,
lianTwistedBilayerGraphene2019,koziiNematicSuperconductivityStabilized2019,
guAntiferromagnetismChiralDwave2019}. 
The integration of monolayer WSe\textsubscript{2} demonstrates the impact of the van der Waals environment and 
proximity effects on the rich phase diagram of TBG. In a broader context, this approach opens the future prospect of 
controlling the range of novel correlated phases available in TBG and similar structures by carefully engineering 
the surrounding layers, and it highlights a key tool for disentangling the mechanisms driving the different correlated states. 

%\end{linenumbers}

%%TC:ignore
\noindent {\bf References:}

%\bibliographystyle{naturemag} %this is the old bibtex style that works well
%\bibliography{TBG_WSe2_btex2}

\noindent {\bf Acknowledgments:} We acknowledge discussions with Hechen Ren, Ding Zhong, Yang Peng, Gil Refael, 
Felix von Oppen, Jim Eisenstein, and Patrick Lee. The device nanofabrication was performed at the Kavli Nanoscience 
Institute (KNI) at Caltech. {\bf Funding:} This work was supported by NSF through
program CAREER DMR-1753306 and grant DMR-1723367, Gist-Caltech memorandum of understanding and the Army 
Research Office under Grant Award W911NF-17-1-0323. Nanofabrication performed by Y.Z. has been 
supported by DOE-QIS program (DE-SC0019166). J.A. and S.N.-P. also acknowledge the support of IQIM 
(NSF funded physics frontiers center). A.T. and J.A. are grateful for support from the Walter Burke 
Institute for Theoretical Physics at Caltech and the Gordon and Betty Moore Foundation’s EPiQS Initiative, 
Grant GBMF8682. The material synthesis at UW was supported as part of Programmable Quantum Materials, an Energy 
Frontier Research Center funded by the U.S. Department of Energy (DOE), Office of Science, Basic Energy Sciences 
(BES), under award DE-SC0019443 and the Gordon and Betty Moore Foundation’s EPiQS Initiative, Grant GBMF6759 to J.-H.C.

\noindent {\bf Author Contribution:} H.A., R.P., Y.Z., and S.N.-P. designed the experiment. 
H.A. made 
the TBG-WSe\textsubscript{2} devices assisted by Y.Z., H.K. and Y.C.  H.A. and R.P., performed the measurements. 
Y.Z. performed measurements on initial TBG devices. H.A., R.P., and S.N.-P. analyzed the data.
A.T. and J.A. developed the continuum model that includes spin-orbit interaction and performed model 
calculations. Z.L., I.Z.W., X.X., and J.-H.C. provided WSe\textsubscript{2} crystals. 
K.W. and T.T. provided hBN crystals. H.A., R.P., Y.Z., A.T., J.A., and S.N.-P. wrote the manuscript with input from 
other authors. S.N.-P. supervised the project. 

\noindent {\bf Data availability:} The data that support the findings of this 
study are available from the corresponding authors on reasonable request.

\clearpage

\clearpage

\clearpage

%supplemenatry 
\beginsupplement
%\section*{Extended Data Figures:}
\begin{figure}[ht]
\label{exfig:fabrication}
\end{figure}

%\clearpage

\begin{figure}[ht]
\label{exfig:opticalimages}
\end{figure}
%\clearpage

\begin{figure}[ht]
\label{exfig:D1extratemp}
\end{figure}

%\clearpage

\begin{figure}[ht]
\label{exfig:D3extra}
\end{figure}

%\clearpage

\begin{figure}[ht]
\label{exfig:D2fraunhofer}
\end{figure}

%\clearpage

\begin{figure}[ht]
\label{exfig:D1fraunhofer}
\end{figure}
%%TC:endignore

\end{document}